\documentclass[acmsmall, nonacm, svgnames]{acmart}

\setcopyright{none}
\acmBooktitle{HATRA '22: Human Aspects of Types and Reasoning Assistants,
              December 5, 2021, Auckland, New Zealand}

\usepackage{booktabs}   %% For formal tables:
\usepackage{fancyvrb}
\usepackage{color}
\usepackage{graphicx}
\usepackage{listings}
\usepackage{subcaption} %% For complex figures with subfigures/subcaptions
\usepackage{ebproof}
\usepackage{amsmath}
\usepackage{latexsym} % this must be included after amssymb and stix, \leadsto
\usepackage{mathtools}
\usepackage{trimclip}
\usepackage{hyperref}

\graphicspath{{./images/}}

\makeatletter
\def\PY@reset{\let\PY@it=\relax \let\PY@bf=\relax%
    \let\PY@ul=\relax \let\PY@tc=\relax%
    \let\PY@bc=\relax \let\PY@ff=\relax}
\def\PY@tok#1{\csname PY@tok@#1\endcsname}
\def\PY@toks#1+{\ifx\relax#1\empty\else%
    \PY@tok{#1}\expandafter\PY@toks\fi}
\def\PY@do#1{\PY@bc{\PY@tc{\PY@ul{%
    \PY@it{\PY@bf{\PY@ff{#1}}}}}}}
\def\PY#1#2{\PY@reset\PY@toks#1+\relax+\PY@do{#2}}

\expandafter\def\csname PY@tok@gd\endcsname{\def\PY@tc##1{\textcolor[rgb]{0.63,0.00,0.00}{##1}}}
\expandafter\def\csname PY@tok@gu\endcsname{\let\PY@bf=\textbf\def\PY@tc##1{\textcolor[rgb]{0.50,0.00,0.50}{##1}}}
\expandafter\def\csname PY@tok@gt\endcsname{\def\PY@tc##1{\textcolor[rgb]{0.00,0.27,0.87}{##1}}}
\expandafter\def\csname PY@tok@gs\endcsname{\let\PY@bf=\textbf}
\expandafter\def\csname PY@tok@gr\endcsname{\def\PY@tc##1{\textcolor[rgb]{1.00,0.00,0.00}{##1}}}
\expandafter\def\csname PY@tok@cm\endcsname{\let\PY@it=\textit\def\PY@tc##1{\textcolor[rgb]{0.25,0.50,0.50}{##1}}}
\expandafter\def\csname PY@tok@vg\endcsname{\def\PY@tc##1{\textcolor[rgb]{0.10,0.09,0.49}{##1}}}
\expandafter\def\csname PY@tok@vi\endcsname{\def\PY@tc##1{\textcolor[rgb]{0.10,0.09,0.49}{##1}}}
\expandafter\def\csname PY@tok@vm\endcsname{\def\PY@tc##1{\textcolor[rgb]{0.10,0.09,0.49}{##1}}}
\expandafter\def\csname PY@tok@mh\endcsname{\def\PY@tc##1{\textcolor[rgb]{0.40,0.40,0.40}{##1}}}
\expandafter\def\csname PY@tok@cs\endcsname{\let\PY@it=\textit\def\PY@tc##1{\textcolor[rgb]{0.25,0.50,0.50}{##1}}}
\expandafter\def\csname PY@tok@ge\endcsname{\let\PY@it=\textit}
\expandafter\def\csname PY@tok@vc\endcsname{\def\PY@tc##1{\textcolor[rgb]{0.10,0.09,0.49}{##1}}}
\expandafter\def\csname PY@tok@il\endcsname{\def\PY@tc##1{\textcolor[rgb]{0.40,0.40,0.40}{##1}}}
\expandafter\def\csname PY@tok@go\endcsname{\def\PY@tc##1{\textcolor[rgb]{0.53,0.53,0.53}{##1}}}
\expandafter\def\csname PY@tok@cp\endcsname{\def\PY@tc##1{\textcolor[rgb]{0.74,0.48,0.00}{##1}}}
\expandafter\def\csname PY@tok@gi\endcsname{\def\PY@tc##1{\textcolor[rgb]{0.00,0.63,0.00}{##1}}}
\expandafter\def\csname PY@tok@gh\endcsname{\let\PY@bf=\textbf\def\PY@tc##1{\textcolor[rgb]{0.00,0.00,0.50}{##1}}}
\expandafter\def\csname PY@tok@ni\endcsname{\let\PY@bf=\textbf\def\PY@tc##1{\textcolor[rgb]{0.60,0.60,0.60}{##1}}}
\expandafter\def\csname PY@tok@nl\endcsname{\def\PY@tc##1{\textcolor[rgb]{0.63,0.63,0.00}{##1}}}
\expandafter\def\csname PY@tok@nn\endcsname{\let\PY@bf=\textbf\def\PY@tc##1{\textcolor[rgb]{0.00,0.00,1.00}{##1}}}
\expandafter\def\csname PY@tok@no\endcsname{\def\PY@tc##1{\textcolor[rgb]{0.53,0.00,0.00}{##1}}}
\expandafter\def\csname PY@tok@na\endcsname{\def\PY@tc##1{\textcolor[rgb]{0.49,0.56,0.16}{##1}}}
\expandafter\def\csname PY@tok@nb\endcsname{\def\PY@tc##1{\textcolor[rgb]{0.00,0.50,0.00}{##1}}}
\expandafter\def\csname PY@tok@nc\endcsname{\let\PY@bf=\textbf\def\PY@tc##1{\textcolor[rgb]{0.00,0.00,1.00}{##1}}}
\expandafter\def\csname PY@tok@nd\endcsname{\def\PY@tc##1{\textcolor[rgb]{0.67,0.13,1.00}{##1}}}
\expandafter\def\csname PY@tok@ne\endcsname{\let\PY@bf=\textbf\def\PY@tc##1{\textcolor[rgb]{0.82,0.25,0.23}{##1}}}
\expandafter\def\csname PY@tok@nf\endcsname{\def\PY@tc##1{\textcolor[rgb]{0.00,0.00,1.00}{##1}}}
\expandafter\def\csname PY@tok@si\endcsname{\let\PY@bf=\textbf\def\PY@tc##1{\textcolor[rgb]{0.73,0.40,0.53}{##1}}}
\expandafter\def\csname PY@tok@s2\endcsname{\def\PY@tc##1{\textcolor[rgb]{0.73,0.13,0.13}{##1}}}
\expandafter\def\csname PY@tok@nt\endcsname{\let\PY@bf=\textbf\def\PY@tc##1{\textcolor[rgb]{0.00,0.50,0.00}{##1}}}
\expandafter\def\csname PY@tok@nv\endcsname{\def\PY@tc##1{\textcolor[rgb]{0.10,0.09,0.49}{##1}}}
\expandafter\def\csname PY@tok@s1\endcsname{\def\PY@tc##1{\textcolor[rgb]{0.73,0.13,0.13}{##1}}}
\expandafter\def\csname PY@tok@dl\endcsname{\def\PY@tc##1{\textcolor[rgb]{0.73,0.13,0.13}{##1}}}
\expandafter\def\csname PY@tok@ch\endcsname{\let\PY@it=\textit\def\PY@tc##1{\textcolor[rgb]{0.25,0.50,0.50}{##1}}}
\expandafter\def\csname PY@tok@m\endcsname{\def\PY@tc##1{\textcolor[rgb]{0.40,0.40,0.40}{##1}}}
\expandafter\def\csname PY@tok@gp\endcsname{\let\PY@bf=\textbf\def\PY@tc##1{\textcolor[rgb]{0.00,0.00,0.50}{##1}}}
\expandafter\def\csname PY@tok@sh\endcsname{\def\PY@tc##1{\textcolor[rgb]{0.73,0.13,0.13}{##1}}}
\expandafter\def\csname PY@tok@ow\endcsname{\let\PY@bf=\textbf\def\PY@tc##1{\textcolor[rgb]{0.67,0.13,1.00}{##1}}}
\expandafter\def\csname PY@tok@sx\endcsname{\def\PY@tc##1{\textcolor[rgb]{0.00,0.50,0.00}{##1}}}
\expandafter\def\csname PY@tok@bp\endcsname{\def\PY@tc##1{\textcolor[rgb]{0.00,0.50,0.00}{##1}}}
\expandafter\def\csname PY@tok@c1\endcsname{\let\PY@it=\textit\def\PY@tc##1{\textcolor[rgb]{0.25,0.50,0.50}{##1}}}
\expandafter\def\csname PY@tok@fm\endcsname{\def\PY@tc##1{\textcolor[rgb]{0.00,0.00,1.00}{##1}}}
\expandafter\def\csname PY@tok@o\endcsname{\def\PY@tc##1{\textcolor[rgb]{0.40,0.40,0.40}{##1}}}
\expandafter\def\csname PY@tok@kc\endcsname{\let\PY@bf=\textbf\def\PY@tc##1{\textcolor[rgb]{0.00,0.50,0.00}{##1}}}
\expandafter\def\csname PY@tok@c\endcsname{\let\PY@it=\textit\def\PY@tc##1{\textcolor[rgb]{0.25,0.50,0.50}{##1}}}
\expandafter\def\csname PY@tok@mf\endcsname{\def\PY@tc##1{\textcolor[rgb]{0.40,0.40,0.40}{##1}}}
\expandafter\def\csname PY@tok@err\endcsname{\def\PY@bc##1{\setlength{\fboxsep}{0pt}\fcolorbox[rgb]{1.00,0.00,0.00}{1,1,1}{\strut ##1}}}
\expandafter\def\csname PY@tok@mb\endcsname{\def\PY@tc##1{\textcolor[rgb]{0.40,0.40,0.40}{##1}}}
\expandafter\def\csname PY@tok@ss\endcsname{\def\PY@tc##1{\textcolor[rgb]{0.10,0.09,0.49}{##1}}}
\expandafter\def\csname PY@tok@sr\endcsname{\def\PY@tc##1{\textcolor[rgb]{0.73,0.40,0.53}{##1}}}
\expandafter\def\csname PY@tok@mo\endcsname{\def\PY@tc##1{\textcolor[rgb]{0.40,0.40,0.40}{##1}}}
\expandafter\def\csname PY@tok@kd\endcsname{\let\PY@bf=\textbf\def\PY@tc##1{\textcolor[rgb]{0.00,0.50,0.00}{##1}}}
\expandafter\def\csname PY@tok@mi\endcsname{\def\PY@tc##1{\textcolor[rgb]{0.40,0.40,0.40}{##1}}}
\expandafter\def\csname PY@tok@kn\endcsname{\let\PY@bf=\textbf\def\PY@tc##1{\textcolor[rgb]{0.00,0.50,0.00}{##1}}}
\expandafter\def\csname PY@tok@cpf\endcsname{\let\PY@it=\textit\def\PY@tc##1{\textcolor[rgb]{0.25,0.50,0.50}{##1}}}
\expandafter\def\csname PY@tok@kr\endcsname{\let\PY@bf=\textbf\def\PY@tc##1{\textcolor[rgb]{0.00,0.50,0.00}{##1}}}
\expandafter\def\csname PY@tok@s\endcsname{\def\PY@tc##1{\textcolor[rgb]{0.73,0.13,0.13}{##1}}}
\expandafter\def\csname PY@tok@kp\endcsname{\def\PY@tc##1{\textcolor[rgb]{0.00,0.50,0.00}{##1}}}
\expandafter\def\csname PY@tok@w\endcsname{\def\PY@tc##1{\textcolor[rgb]{0.73,0.73,0.73}{##1}}}
\expandafter\def\csname PY@tok@kt\endcsname{\def\PY@tc##1{\textcolor[rgb]{0.69,0.00,0.25}{##1}}}
\expandafter\def\csname PY@tok@sc\endcsname{\def\PY@tc##1{\textcolor[rgb]{0.73,0.13,0.13}{##1}}}
\expandafter\def\csname PY@tok@sb\endcsname{\def\PY@tc##1{\textcolor[rgb]{0.73,0.13,0.13}{##1}}}
\expandafter\def\csname PY@tok@sa\endcsname{\def\PY@tc##1{\textcolor[rgb]{0.73,0.13,0.13}{##1}}}
\expandafter\def\csname PY@tok@k\endcsname{\let\PY@bf=\textbf\def\PY@tc##1{\textcolor[rgb]{0.00,0.50,0.00}{##1}}}
\expandafter\def\csname PY@tok@se\endcsname{\let\PY@bf=\textbf\def\PY@tc##1{\textcolor[rgb]{0.73,0.40,0.13}{##1}}}
\expandafter\def\csname PY@tok@sd\endcsname{\let\PY@it=\textit\def\PY@tc##1{\textcolor[rgb]{0.73,0.13,0.13}{##1}}}

\def\PYZus{\char`\_}
\def\PYZob{\char`\{}
\def\PYZcb{\char`\}}

\def\PYZam{\char`\&}
\def\PYZlt{\char`\<}
\def\PYZgt{\char`\>}

\def\PYZhy{\char`\-}

\def\PYZdq{\char`\"}

\makeatother

\newcommand{\allow}{\PY{k}{\texttt{allow}}}
\newcommand{\flow}[1]{\PY{k}{\texttt{flow}}\texttt{ #1}}
\newcommand{\textmod}{\PY{k}{\texttt{mod}}}
\newcommand{\textlet}{\PY{k}{\texttt{let}}}
\newcommand{\self}{\PY{n+nb+bp}{\texttt{self}}}
\newcommand{\textif}{\PY{k}{\texttt{if}}}
\newcommand{\withflow}{\PY{k}{\texttt{with flow}}}
\newcommand{\either}{\PY{n+nc}{\texttt{Either}}}
\newcommand{\result}{\PY{n+nc}{\texttt{Result}}}

\newcommand{\new}[1][brown]{\textcolor{#1}}

\newcommand{\rvdash}{~\reflectbox{$\vdash$}~}
\newcommand{\plustriangle}{+\!\!>}

\newcommand{\curlybr}[1]{\{~#1~\}}

\newcommand{\typingrule}[7][]{
    \begin{prooftree}
        #4
        \infer#3[#1{\textsc{#2}}]{
            #5 \vdash{} #6 \Rightarrow{} #7
        }
    \end{prooftree}
}

\newcommand{\typejgmt}[3][\Sigma; \Delta; \Gamma; \Theta;
    \new{\Pi; \Psi; \Xi}] {
    #1 \vdash{} #2 \Rightarrow{} #3
}

\newcommand{\freeregions}{\operatorname{free\text{-}regions}}
\newcommand{\foldr}{\operatorname{foldr}}
\newcommand{\gcloans}{\operatorname{gc\text{-}loans}}

\newcommand{\allowed}{\operatorname{is\text{-}allowed}}
\newcommand{\getperms}{\operatorname{get\text{-}min\text{-}perms}}

\newcommand{\deltaplace}{\operatorname{\delta\text{-}place}}

\newcommand{\leaves}{\operatorname{leaves}}

\newcommand{\assigndeps}{\operatorname{assign\text{-}deps}}

\newcommand{\deltaleaves}{\operatorname{\delta\text{-}leaves}}

\newcommand{\sxsc}{\mathrm{SX}}
\newcommand{\xisc}{\mathrm{XI}}
\newcommand{\sisc}{\mathrm{SI}}
\newcommand{\sdsc}{\mathrm{SD}}

\newlength{\trianglewidth}
\newcommand{\righttrianglebar}{%
    \mathrel{\makebox[\trianglewidth]{%
        \makebox[\trianglewidth]{\(-\)}%
        \hspace*{-\trianglewidth}%
        \makebox[\trianglewidth]{\(\vartriangleright\)}%
    }}%
}

\newcommand{\uniq}{\texttt{uniq}}

\begin{document}

%% Title information
\title{Static Information Flow Control Made Simpler}
%% [Short Title] is optional;
%% when present, will be used in
%% header instead of Full Title.
% \titlenote{with title note}           %% \titlenote is optional;
%% can be repeated if necessary;
%% contents suppressed with 'anonymous'
% \subtitle{Subtitle}                   %% \subtitle is optional
% \subtitlenote{with subtitle note}     %% \subtitlenote is optional;
%% can be repeated if necessary;
%% contents suppressed with 'anonymous'

%% Author information
%% Contents and number of authors suppressed with 'anonymous'.
%% Each author should be introduced by \author, followed by
%% \authornote (optional), \orcid (optional), \affiliation, and
%% \email.
%% An author may have multiple affiliations and/or emails; repeat the
%% appropriate command.
%% Many elements are not rendered, but should be provided for metadata
%% extraction tools.

\author{Hemant Gouni}
\email{gouni008@umn.edu}          %% \email is recommended
\affiliation{
    \institution{University of Minnesota}            %% \institution is required
    \city{Minneapolis}
    \state{Minnesota}
    \country{USA}                    %% \country is recommended
}

\author{Jonathan Aldrich}
\email{jonathan.aldrich@cs.cmu.edu}
\affiliation{
  \institution{Carnegie Mellon University}
  \city{Pittsburgh}
  \state{Pennsylvania}
  \country{USA}
}

%% Abstract
%% Note: \begin{abstract}...\end{abstract} environment must come
%% before \maketitle command
\begin{abstract}
    Static information flow control (IFC) systems provide the ability to
    restrict data flows within a program, enabling vulnerable functionality or
    confidential data to be statically isolated from unsecured data or program
    logic. Despite the wide applicability of IFC as a mechanism for
    guaranteeing \textit{confidentiality} and \textit{integrity}--- the
    fundamental properties on which computer security relies--- existing IFC
    systems have seen little use, requiring users to reason about complicated
    mechanisms such as lattices of security labels and dual notions of
    confidentiality and integrity within these lattices. We propose a system
    that diverges significantly from previous work on information flow control,
    opting to reason \textit{directly} about the data that programmers already
    work with. In doing so, we naturally and seamlessly combine the clasically
    separate notions of confidentiality and integrity into one unified
    framework, further simplifying reasoning. We motivate and showcase our work
    through two case studies on TLS private key management: one for Rocket, a
    popular Rust web framework, and another for Conduit, a server
    implementation for the Matrix messaging service written in Rust.
\end{abstract}

\begin{CCSXML}
<ccs2012>
   <concept>
       <concept_id>10011007.10011074.10011099.10011692</concept_id>
       <concept_desc>Software and its engineering~Formal software verification</concept_desc>
       <concept_significance>500</concept_significance>
       </concept>
   <concept>
       <concept_id>10011007.10011006.10011060.10011690</concept_id>
       <concept_desc>Software and its engineering~Specification languages</concept_desc>
       <concept_significance>500</concept_significance>
       </concept>
 </ccs2012>
\end{CCSXML}

\ccsdesc[500]{Software and its engineering~Formal software verification}
\ccsdesc[500]{Software and its engineering~Specification languages}

%% End of generated code

%% Keywords
%% comma separated list
\keywords{Information Flow Control, Security Type Systems, Accessible Correctness}  %% \keywords are mandatory in final camera-ready submission

%% \maketitle
%% Note: \maketitle command must come after title commands, author
%% commands, abstract environment, Computing Classification System
%% environment and commands, and keywords command.
\maketitle

\section{Introduction}

\begin{figure}[h]
    \includegraphics[scale=0.7]{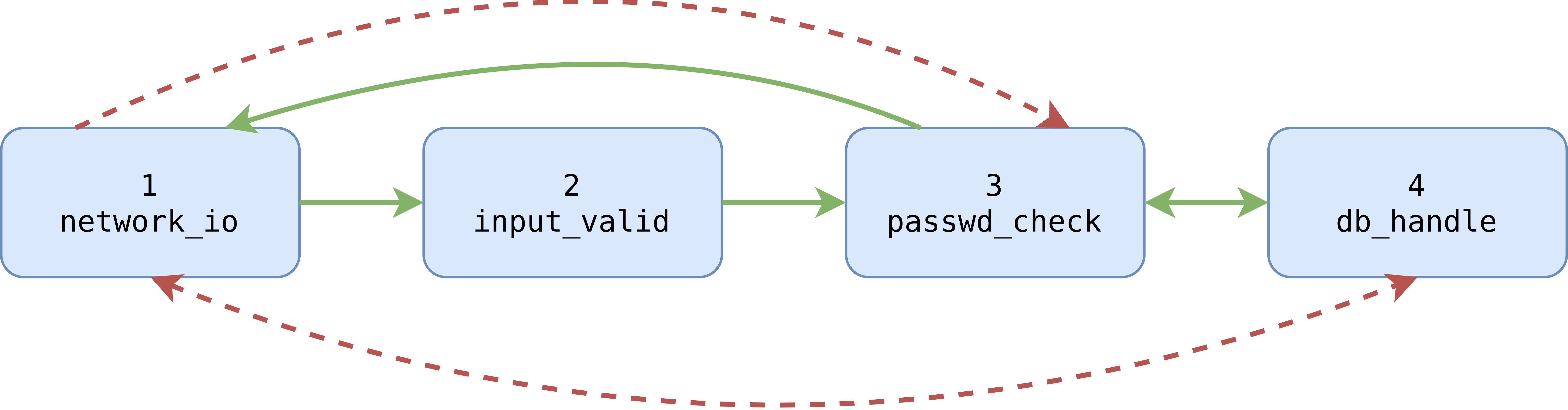}
    \caption{Permitted flows in our example program}\label{fig:flow_diagram}
\end{figure}

Security is a paramount concern in a pervasively digital world. The vast
increase in digitization of the past decade has come with a proportional
increase in the severity of data breaches~\cite{warren2016cybercrime},
warranting renewed research into accessible, yet powerful, methods of declaring
and enforcing security specifications. Information flow control is a program
analysis technique that aims to prevent undesirable data flows. For example, in
a program implementing a simple web application for performing authentication,
there may be modules for (1) performing network I/O, (2) validating user input,
(3) checking passwords, and (4) interacting with a database. A reasonable
information flow specification, shown in Fig.~\ref{fig:flow_diagram}, might be
that (1) should never flow to (3) or (4), and that data from (4) should never
flow to (1). Green (solid) arrows in the diagram denote the flows
\textit{required for the program to function}, and red (dashed) arrows denote
the flows which must be \textit{prevented to ensure the program's security}. A
\textit{flow} from, for instance, (1) to (4), occurs when data from (1), or
which is transitively dependent on (1), flows to (4). Note that certain flows
in the diagram, such as from (1) to (4), are denied directly but permitted
transitively--- specifically, after data from (1) passes through (2). This is
addressed during the case study in Section~\ref{sec:motivation}.

A classical information flow control system~\cite{10.1145/292540.292561,
simonet2003flow} might support the specification of the flow architecture in
Fig.~\ref{fig:flow_diagram} by providing two lattices: one for reasoning about
confidentiality, or the property that confidential data does not leak to
inappropriately visible places, and one for integrity, or the property that
untrusted data does not propagate to inappropriately integral places. All data
in the program must have a label relevant to each lattice, denoting its levels
of confidentiality and integrity. Two separate lattices are required for each
property owing to the duality between them: two distinct and opposing
invariants must be enforced when data is labeled as both highly confidential
and highly integral, such as with \texttt{db\_handle} in
Fig.~\ref{fig:flow_diagram}. First, it must never flow to data labeled as being
of low confidentiality, such as \texttt{network\_io}. Second, data labeled as
being of low integrity, again like \texttt{network\_io}, must never flow to
\texttt{db\_handle}. This separate treatment complicates reasoning--- for
example, Jif~\cite{DBLP:journals/corr/Pullicino14}, a state of the art IFC
system for Java, provides entirely new (dual) semantics for lattice operations
when reasoning about integrity versus confidentiality.

Our system resolves this issue by presenting a single, straightforward
abstraction for declaring information flow policies, simplifying specifications
while retaining expressivity. Instead of organizing data inside a lattice and
reasoning about flows in the context of it, we opt to allow programmers to
directly specify which flows between places in a program are not permitted, as
presented in the architecture in Fig.~\ref{fig:flow_diagram}. A declaration
\flow{b ->! c} states that data affected by variable \texttt{b} cannot flow to
variable \texttt{c}, and a declaration \flow{\textmod{} network\_io ->!
\textmod{} db\_handle} states that data affected by the \texttt{network\_io}
module cannot flow to the \texttt{db\_handle} module. We will refer to the part
before the arrow as the \textit{source}, and the part after as the
\textit{destination}. This streamlined annotation language contributes to a
system which is:

% TODO: make the diagram on page 1 more color blind friendly

\begin{enumerate}
    \item \textbf{Uniform:} a flow annotation \flow{b ->! c} may be an
          integrity specification, a confidentiality specification, or
          potentially both--- the programmer need not shift their reasoning
    \item \textbf{Incremental:} our approach to flow annotations encourages
          \textit{partial} and \textit{local} specifications on the way to
          proving higher level program properties, and
    \item \textbf{Simple:} we provide a small number of straightforward
          constructs which are highly general and composable, minimizing changes
          needed to the semantics of the language and easing adoption of the
          system by new users.
\end{enumerate}

We will explore these properties further by analyzing cryptographic key
management in two popular Rust programs, Rocket and Conduit. A brief formal
description of our IFC system, as an extension to the Rust type system, is
provided in the appendix for the curious reader.

\section{Motivation: Managing Keys}\label{sec:motivation}

\subsection{Case Study: Rocket}\label{subsec:case_study_rocket}

\begin{figure}[h]
    \input{caseStudyRocket}
    \caption{Verifying non-leakage of TLS key data in Rocket}\label{fig:case_study_rocket}
\end{figure}

\href{https://github.com/SergioBenitez/Rocket}{Rocket} is a fully fledged and
widely used HTTP framework for Rust. Like many web frameworks, it provides
optional support for serving traffic over Transport Layer Security, or TLS,
connections. TLS is an essential feature for securing data between clients,
such as browsers, and servers, encrypting data in transit. A critical component
of TLS is a public-private keypair held by the server, in this case Rocket.
Each domain which aims to communicate securely with web browsers and clients at
large must obtain a certified keypair, which is valid for \textit{any} traffic
served by it. While a typical web service consists of a large
number of physical servers, each of them may use the same TLS keypair, assuming
they are serving a single domain. This is convenient, but risky--- leakage of
the private key for a domain from any of its servers risks compromising the
security of the entire service. It is essential, then, that this
does not occur, and an IFC system may be able to provide that assurance.

Most of our discussion will be directed at the particular Rocket code path that
deals with conditionally configuring TLS functionality, shown in
Fig.~\ref{fig:case_study_rocket}. Specifically, we analyze the function
\texttt{default\_tcp\_http\_server} in file \texttt{core/lib/src/server.rs},
on commit \texttt{2fc4b156} of the Rocket git repository. In order to render
this section of code amenable to information flow verification, some minimal
refactoring was required; we will return to this later. We focus our attention
here because it is the primary part of the codebase where key data is accessed
directly (through \texttt{to\_native\_config} and \texttt{TlsListener::bind} on
lines 13 and 19), increasing the potential for leaks and need for verification.
Verification of the latter two functions is not as interesting, because their
implementation is largely straightforward or reliant on primitives provided by
Rustls, a TLS library for Rust. A summary of the displayed Rocket code follows.

Ignoring the information flow constructs, the code in Fig.
~\ref{fig:case_study_rocket}:

\begin{itemize}
    \item On lines 6 and 7: Binds a variable, \texttt{listener}, which
          initially holds a TCP socket
    \item On lines 9 and 10: Checks that TLS is enabled and that a TLS
          configuration is provided
    \item In lines 13 through 19: Extracts the provided TLS configuration into
          a structure understood by \texttt{TlsListener::bind} and calls it,
          setting \texttt{listener} to the returned TLS socket.
\end{itemize}

We now discuss the IFC specifications for \texttt{\self.config.tls.key}, which
contains key data, in Fig.~\ref{fig:case_study_rocket}. The flow declaration on
the first line prohibits the private key from flowing transitively to any
functions which could perform I/O operations, such as those in
\texttt{std::io}, \texttt{std::fs}, \texttt{std::net}, \texttt{std::ffi}, and
\texttt{std::os}. The same property could be expressed as a series of
individual flow declarations on each relevant namespace, but the use of a macro
here provides a convenient abstraction. The next declaration employs a special
place understood by flow rules: \texttt{*}, which applies to all variables,
along with the return value. This ensures that key data does not leak to any
place without being explicitly permitted to do so. We care about flows to
places, instead of merely flows to I/O operations, because of abstraction: a
module may never perform I/O internally, but it may communicate with another
that does, and verifying that it does so safely requires assertions about
information flow inside the module. Permission for flows may be granted in two
ways: through \textit{overriding} the flow rules we have already declared, or
by using the \allow{} construct.

Overriding occurs whenever a rule is declared that is more specific than
another. Specificity is determined by the length of the prefix being accessed
within a variable or namespace. For example, given a rule \flow{b ->! c} in
some scope, where \texttt{b} is a struct with field \texttt{f}, a rule
\flow{b.f -> c} declared in that scope would act as an override, permitting
flows from \texttt{b.f} to \texttt{c} (but not from any other fields of
\texttt{b}). Additionally, there are two invariants enforced by the system on
flow declarations, which ensure that it can always find a \textit{maximally
specific} rule for any given flow. First, it is an error to declare
contradictory flow rules. Second, in the case of two opposing flow rules where
one flow rule has a more specific destination than another, and the other a
more specific source, the one that denies the flow is chosen. This might occur
in the case of two flow rules \flow{a.b ->! c} and \flow{a -> c.d}--- a flow
from \texttt{a.b} to \texttt{c.d} would be denied because the first rule is
denies the flow. This is done for safety: denied flows are visible, but
permitted flows are not. We can now analyze the last three flow rules in
Fig.~\ref{fig:case_study_rocket}.

Starting with the rule on line 11, we see slightly different syntax,
\withflow{}, in the declaration. This is done because we cannot declare the
flow rule on \texttt{config} before it has been bound, and by the time the
binding happens, it is already too late, and a violation of the rule on line 2
is inevitable. \withflow{} is syntactic sugar provided for the \textlet{}
construct which inserts a flow declaration between binding and initialization.
We proceed similarly for the flow rules on lines 14 and 16, overriding the flow
rule on line 2 as needed, but another element bears discussion: function calls.
When a function is called with data affected by rules denying flows, the
information flow control system checks that the flow policies declared by the
function are compatible with those from the calling context. Only
mutably referenced arguments, the return value, and any restrictions on flows
to functions are checked here; the precise mechanics are discussed further
in Appendix~\ref{sec:formalizing}. Importantly, we only look at the publicly
visible portion of a function's signature, preserving abstraction and ensuring
modularity. We are able to treat functions in this way because Rust programs
generally do not make use of global mutable state--- treating it as an error,
then, does not result in a materially more conservative system. As a result,
for functions which do not perform IO, we must only consider their arguments
and return value. In this case, the call to \texttt{to\_native\_config}
receives an immutable reference and does not take any other arguments, so the
\withflow{} declaration suffices to allow the call. The function call on line
19, which takes its two arguments by move, passes similarly. Note that, having
explicitly permitted flows from \texttt{\self.config.tls.key} to
\texttt{config}, \texttt{conf}, and \texttt{listener}, flows from these four
variables to any other variables will now cause an error.

% you can't move something that has mutable references inside it, so do
% we need the function call judgement to be so complicated?
%
% well you can move the actual reference though?

% non restrictive relative clause!?
Lastly, we arrive at the assignment to \texttt{listener} on the second-to-last
line. \allow{} evaluates an expression and returns the result, but without any
of the expression's dependencies. It is comparable to other information flow
systems' \texttt{declassify} (for confidentiality) and \texttt{endorse} (for
integrity) constructs~\cite{hedin2012perspective} in providing an explicit
escape hatch, but generalizes over both. In this case, we use \allow{} to
remove \texttt{listener}'s dependency on \texttt{\self.config.tls.key},
assigning it to itself so that further uses will not produce errors. The usage
of \allow{} here is safe because, importantly, \texttt{TlsListener} no longer
provides access to raw key data, only to signing operations with the key. This
is still a concern, but a much lesser one--- data dependencies on
\texttt{listener} do not risk revealing information about the cryptographic
key. Further, it is reasonable to wonder why we must use \allow{} on
\texttt{listener}, outside the scope of the preceding conditionals, instead of
on, say, \texttt{conf} in the argument list to \texttt{TlsListener::bind}.
\textit{Control flow dependencies} are what make this necessary. These are
\textit{implicit} flows that take place when \textit{explicit} flows occur
inside a conditional branch. In Fig.~\ref{fig:case_study_rocket}, this
crucially happens when \texttt{listener} is assigned to the result of
evaluating \texttt{TlsListener::bind}, where it gains a dependency on
\texttt{\self.config.tls.key} not only because of the function's return value,
but also because the guards for both \textif{} statements, on lines 9 and 10,
are dependent on \texttt{\self.config.tls.key}. As a result, applying \allow{}
inside the conditional would remove the dependency from \texttt{bind}'s return
value, but not from the conditional branches, which would require two
additional usages on the guards themselves. Instead, we do so once outside both
conditionals and reassign to \texttt{listener}.

Control flow dependencies also explain why a minor refactor was necessary in
order to ease information flow reasoning. \texttt{default\_tcp\_http\_server},
as present in the Rocket repository at the referenced commit, unnecessarily
includes several assignments inside the conditionals, in addition to logic
afterwards which depends on those assignments. Moving all of this functionality
outside of the conditional branches removes unneeded dependencies on the value
of \texttt{\self.config.tls.key} and additionally results in less overall code
duplication. In general, we believe that the introduction of an information
flow control type system to programs which did not have it before will
encourage shifts in coding style, just as typed programs differ from untyped
programs in order to better cooperate with (and take better advantage of) the
type system.

To summarize, we show that, under our formal system, we can concisely prove a
strong information flow property about TLS key management within Rocket,
specifically, that no TLS key data leaks from
\texttt{default\_tcp\_http\_server}. Though bugs were not identified while
demonstrating our theory of information flow, note that our specifications
would suffice to prevent the occurrence of \textit{future} bugs within
relatively complicated and error-prone code. Moreover, this property is
\textit{local}, only concerning the function itself and two of its callees.
Specifications beyond those discussed here could be used to prove higher level
properties about the absence of key leakage from the entirety of Rocket. We now
apply the mechanisms we have discussed here to a second program.

% how do we deal with flow policies declared on functions (like flows to
% functions)
%
% they can be modeled the same as dereferences! think about the function as a
% 'place' to which we must prevent flows, passing every function when calling

\subsection{Case Study: Conduit}\label{subsec:case_study_conduit}

\href{https://gitlab.com/famedly/conduit}{Conduit} is a Rust implementation of
the server portion of the \href{https://matrix.org}{Matrix} chat protocol. As a
federated messaging service (meaning there does not exist a single canonical
server, but a cooperating network of them, much like email), Matrix relies
heavily on cryptography to secure and authenticate message contents. This
includes communication between Matrix servers themselves. We analyze a similar
situation to the one identified with Rocket, in which complicated application
initialization logic executes within a context that contains raw key data, and
so might benefit from the use of IFC to explicitly enforce security policies.
We specifically analyze the \texttt{load} function in file
\texttt{src/database/globals.rs}, on commit \texttt{566dc0a} of the Conduit git
repository. The relevant snippet of code is shown in
Fig.~\ref{fig:case_study_conduit1}. As before, we begin with a brief summary of
its functionality.

\begin{figure}
    \include{caseStudyConduit1}
    \caption{Verifying that Conduit does not leak Matrix server keys}\label{fig:case_study_conduit1}
\end{figure}

\begin{itemize}
    \item Lines 6 and 7: Process the raw bytes representing the key into a
        version indicator and a key
    \item Line 9: Perform several steps of validation on the version string and
        key data
    \item Lines 10 through 15: Pass the validated data to a library function
        that handles cryptography for Matrix applications
\end{itemize}

Addressing the information flow constructs in
Fig.~\ref{fig:case_study_conduit1}, we start with the same two declarations as
for Rocket. The first, on line 3, prohibits flows to I/O functions, which might
allow sensitive information to escape, and the second prohibits flows to any
variables, along with the return value of the current function. We allow the
flow on line 7 using the \withflow{} construct and our notion of overriding
flow policies. Overriding occurs here because \texttt{parts} is more specific
than \texttt{*}. Note, again, that flows from \texttt{parts} will still cause
errors.

A similar \withflow{} declaration is made on line 15, again using overriding.
However, lines 10 through 12, immediately prior to it, warrant further
discussion due to their use of closures. The treatment of closures--- and
higher order functions in general--- is straightforward in our system. Besides
the checks done for function calls, discussed in
Section~\ref{subsec:case_study_rocket}, we must perform two additional checks
involving any data captured by the closure. First, the captured data is added
to the set of dependencies which will flow through the closure's return value,
or to any mutable references passed as arguments to it. Second, we ensure that
the closure is never passed any information that might inappropriately flow to
variables captured by it. The flow policy for the current scope remains active
for the body of the closure itself, so any potential flows between captured
variables will be caught when the closure is declared. Luckily, however, the
usages on line 10 and 12 do not need these considerations. The closure
declaration on line 10 does not close over anything, only taking two arguments,
neither of which is a mutable reference, and returning key data, wrapped in a
\result{} type. The closure on line 12 is similar, taking one argument
(which contains an error value, if applicable) and returning another \result{} to
\texttt{keypair}. Accordingly, the \withflow{} declaration on line 15 suffices
to allow \texttt{keypair\_bytes} to flow to \texttt{keypair}.

The function call on line 11 securely abstracts away raw key data, rendering it
inaccessible, just as was the case with Rocket. It is again safe, then, to use
\allow{} on \texttt{keypair}, assigning it to itself to prevent further
information flow violations from uses of it. Having analyzed this snippet from
the Conduit codebase, we note that it is highly similar to the Rocket code
before it, both in terms of the information flow control issues raised and the
simple, declarative solutions offered by our system. We expect this general
pattern of information flow control to arise consistently when managing
sensitive data within application configuration and setup logic.

We introduce one final information flow pattern that arises in the Conduit
codebase as a consequence of the pervasive use of cryptographic operations in
the Matrix protocol, and discuss a natural method of addressing it offered by
our system. This time, the data in question does \textit{not} grant direct
access to the private key, only to cryptographic operations with it. It does,
however, exist alongside application logic that interacts with user input over
the network, so IFC is once again warranted. Fig.~\ref{fig:case_study_conduit2}
shows part of the implementation of \texttt{create\_invite\_route} in file
\texttt{src/server\_server.rs} on the same commit. We start, as usual, with a
short summary of the code.

\begin{figure}
    \include{caseStudyConduit2}
    \caption{The event signing pattern in Conduit}\label{fig:case_study_conduit2}
\end{figure}

\begin{itemize}
    \item Lines 8 through 10: Create a JSON object with a
        \href{https://datatracker.ietf.org/doc/html/rfc8785}{reproducible
        serialized representation} to prepare it to be hashed, assigning it to
        \texttt{signed\_event}
    \item Lines 12 through 15: Hash and sign the JSON object by passing it,
        \texttt{db.globals.keypair()}, and a mutable reference to  % chktex 36
        \texttt{signed\_event} to a library function responsible for handling
        cryptography; \texttt{signed\_event} is assigned the result through the
        reference
\end{itemize}

Preceding and following the code in Fig.~\ref{fig:case_study_conduit2} is a
large amount of logic for inviting a user to a room through the Matrix
protocol--- \texttt{create\_invite\_route} implements the relevant API endpoint
for doing so. This pattern arises not just in this function, but throughout
Conduit, for \textit{every API endpoint which must use key data}. In other
words, large portions of Conduit both have access to
\texttt{db.globals.keypair()}, which is passed to every function that needs it, % chktex 36
and process networked user input. Our IFC system adapts naturally
to this challenge because of its focus on \textit{local} reasoning.

We begin with the usual flow declarations on lines 3 and 4, restricting
\texttt{db.globals.keypair()} from flowing to any functions which perform I/O, % chktex 36
any variables, or the return value. It should be noted that while
\texttt{db.globals.keypair()} cannot reveal raw key data, it may still produce % chktex 36
sensitive information (such as signed events), so data dependent on it should
be prevented from flowing to I/O. We use a single \allow{} in the call to
\texttt{ruma::signatures::hash\_and\_sign\_event} on line 12, because
\texttt{signed\_event} does not pose any risk of compromising access to the key.

Though this is an exceptionally simple flow specification, note that we have
also verified an exceptionally strong information flow property. Besides
\texttt{signed\_event}, no other data in the program is permitted to invoke, or
use data from invoking, \texttt{db.globals.keypair()}. Program logic not
pictured in Figs.~\ref{fig:case_study_conduit1} and
~\ref{fig:case_study_conduit2} for brevity requires further specifications in
order to pass information flow checking (in particular, functions to which
\texttt{db} is passed, whose flow policies would need to be updated in order
for the call to succeed), but the added complexity is negligible.

\subsection{What about lattices?}

We now briefly compare our system to one implementing a generic lattice-based
theory of information flow. We avoid complicated constructs such as
\textit{principals}~\cite{myers2000protecting} and \textit{robust
declassification}~\cite{myers2004enforcing} for which we currently lack
meaningful comparisons.

The most apparent advantage of our system, relative to any that employs
lattices, is that our specifications act as a \textit{declarative} description
of the desired information flow policies in effect for any program scope. At a
glance, it is trivially easy to determine the desired flow architecture, what
data is critical, and how to work within programs annotated with IFC
specifications. Should the user desire to enforce, taking
Fig.~\ref{fig:case_study_rocket} for example, that
\texttt{\self.config.tls.key} does not leak to unsecured areas in the program,
it's necessary only to declare precisely that property, explicitly permitting flows where
needed and explicitly disengaging the IFC system once the language's normal
abstraction mechanisms have guaranteed the property at hand. A lattice-based
system would need to talk about the property that the key does not leak
\textit{indirectly}, giving it a label that indicates high confidentiality and
providing other data with labels of lower confidentiality. Adding additional
flow restrictions, especially integrity policies, would make the desired
information flow architecture much less clear in this setting.

Finally, a last point should be noted about label-based reasoning versus the
direct usage of program places: locality. Generally, our system's ability to
reason about programs locally is not due to any increase in formal reasoning
power over prior work, but can instead be owed to its removal of labels as a
mechanism for abstracting over places in a program--- labels naturally
encourage the creation of specifications that attempt to talk about multiple,
potentially disparate, places at once, instead of incrementally declaring flow
policies on critical data and proving them. For instance, when using a system
such as Flow Caml~\cite{simonet2003flow}, a lattice of security levels is
declared \textit{globally} and enforced for the whole program at once.

% TODO: discuss where the sensitive key data comes into play in all examples?
%
% (it's passed to the function as an argument)

% TODO: Result and Either need syntax highlighting

% Closures:
% same as functions, except we need to keep track of anything the closure
% closes over, and add the dependencies of those things to its return value and
% any mutable references passed to it

% TODO: We have to reformulate get-minimal-permissions a bit!

% See Jonathan's overleaf comment on choosing the deny rule instead of the rule
% with the more specific source

\section{Conclusion and Related Work}

Much related work on information flow control, particularly \textit{usable}
information flow control, includes support for dynamic verification when static
reasoning presents too high a barrier--- \cite{10.1145/292540.292561} does just
this. The formal system described here, however, carries enough novelty in the
fully static realm to warrant an exclusive focus; future work may introduce
dynamic elements.~\cite{simonet2003flow} increases the usability of its IFC
system by precisely \textit{inferring} security labels, which may similarly be
an interesting direction for future work. \cite{10.1145/3519939.3523445}
provides a modular information flow analysis for Rust programs on which we
build our formalism, but it does not provide a way to specify intended
restrictions on information flow.

We have reviewed classical systems of information flow, introducing a
simplified specification language and system constructed around it which we
hope will aid in advancing the accessibility of information flow. We conducted
two case studies in which we demonstrated the natural applicability of our system
to \textit{local} reasoning. After proving the soundness of our system, we hope to
pursue a practical implementation for use on real-world Rust programs.

% Oxide supporting EITHER means we can't accurately estimate...
%
% actually maybe we can! by just treating Either as a boundary! after
% that we view it as a leaf. and i don't see how you can stack allocate
% potentially infinite structures any other way or at least types that
% look infinite sooooo
%
% also we could (should) do the same leaf behavior for type constructors?

% NOTE that it's okay for base types to have \varnothing as their output, bc
% they have no places defined on them

% note that 'which' usually comes after a comma!

% discuss that restructuring code is analogous to typed vs untyped, for the
% refactor. an information flow type system changes what code is idiomatic

% explain transitive flows (why are they okay?)

% the second to last assignment to listener does NOT need to use `with flow`

%% Acknowledgments
\begin{acks}                            %% acks environment is optional
    %% contents suppressed with 'anonymous'
    %% Commands \grantsponsor{<sponsorID>}{<name>}{<url>} and
    %% \grantnum[<url>]{<sponsorID>}{<number>} should be used to
    %% acknowledge financial support and will be used by metadata
    %% extraction tools.
    This research was sponsored by Department of Defense award H9823018D0008
    and National Science Foundation award CCF-1901033. Any opinions, findings,
    and conclusions or recommendations expressed in this material are those of
    the author and do not necessarily reflect the views of the Department of
    Defense or the National Science Foundation.
\end{acks}

%% Bibliography
\bibliography{hatra}


\begin{thebibliography}{9}

%%% the Web address.

\ifx \showCODEN    \undefined \def \showCODEN     #1{\unskip}     \fi
\ifx \showDOI      \undefined \def \showDOI       #1{#1}\fi
\ifx \showISBNx    \undefined \def \showISBNx     #1{\unskip}     \fi
\ifx \showISBNxiii \undefined \def \showISBNxiii  #1{\unskip}     \fi
\ifx \showISSN     \undefined \def \showISSN      #1{\unskip}     \fi
\ifx \showLCCN     \undefined \def \showLCCN      #1{\unskip}     \fi
\ifx \shownote     \undefined \def \shownote      #1{#1}          \fi
\ifx \showarticletitle \undefined \def \showarticletitle #1{#1}   \fi
\ifx \showURL      \undefined \def \showURL       {\relax}        \fi
% invisible to TeX
\providecommand\bibfield[2]{#2}
\providecommand\bibinfo[2]{#2}
\providecommand\natexlab[1]{#1}
\providecommand\showeprint[2][]{arXiv:#2}

\bibitem[Crichton et~al\mbox{.}(2022)]%
        {10.1145/3519939.3523445}
\bibfield{author}{\bibinfo{person}{Will Crichton}, \bibinfo{person}{Marco
  Patrignani}, \bibinfo{person}{Maneesh Agrawala}, {and} \bibinfo{person}{Pat
  Hanrahan}.} \bibinfo{year}{2022}\natexlab{}.
\newblock \showarticletitle{Modular Information Flow through Ownership}. In
  \bibinfo{booktitle}{\emph{Proceedings of the 43rd ACM SIGPLAN International
  Conference on Programming Language Design and Implementation}} (San Diego,
  CA, USA) \emph{(\bibinfo{series}{PLDI 2022})}.
  \bibinfo{publisher}{Association for Computing Machinery},
  \bibinfo{address}{New York, NY, USA}, \bibinfo{pages}{1–14}.
\newblock
\showISBNx{9781450392655}
\urldef\tempurl%
\url{https://doi.org/10.1145/3519939.3523445}
\showDOI{\tempurl}


\bibitem[Hedin and Sabelfeld(2012)]%
        {hedin2012perspective}
\bibfield{author}{\bibinfo{person}{Daniel Hedin} {and} \bibinfo{person}{Andrei
  Sabelfeld}.} \bibinfo{year}{2012}\natexlab{}.
\newblock \showarticletitle{A perspective on information-flow control}.
\newblock In \bibinfo{booktitle}{\emph{Software safety and security}}.
  \bibinfo{publisher}{IOS Press}, \bibinfo{pages}{319--347}.
\newblock


\bibitem[Myers(1999)]%
        {10.1145/292540.292561}
\bibfield{author}{\bibinfo{person}{Andrew~C. Myers}.}
  \bibinfo{year}{1999}\natexlab{}.
\newblock \showarticletitle{JFlow: Practical Mostly-Static Information Flow
  Control}. In \bibinfo{booktitle}{\emph{Proceedings of the 26th ACM
  SIGPLAN-SIGACT Symposium on Principles of Programming Languages}} (San
  Antonio, Texas, USA) \emph{(\bibinfo{series}{POPL '99})}.
  \bibinfo{publisher}{Association for Computing Machinery},
  \bibinfo{address}{New York, NY, USA}, \bibinfo{pages}{228–241}.
\newblock
\showISBNx{1581130953}
\urldef\tempurl%
\url{https://doi.org/10.1145/292540.292561}
\showDOI{\tempurl}


\bibitem[Myers and Liskov(2000)]%
        {myers2000protecting}
\bibfield{author}{\bibinfo{person}{Andrew~C Myers} {and}
  \bibinfo{person}{Barbara Liskov}.} \bibinfo{year}{2000}\natexlab{}.
\newblock \showarticletitle{Protecting privacy using the decentralized label
  model}.
\newblock \bibinfo{journal}{\emph{ACM Transactions on Software Engineering and
  Methodology (TOSEM)}} \bibinfo{volume}{9}, \bibinfo{number}{4}
  (\bibinfo{year}{2000}), \bibinfo{pages}{410--442}.
\newblock


\bibitem[Myers et~al\mbox{.}(2004)]%
        {myers2004enforcing}
\bibfield{author}{\bibinfo{person}{Andrew~C Myers}, \bibinfo{person}{Andrei
  Sabelfeld}, {and} \bibinfo{person}{Steve Zdancewic}.}
  \bibinfo{year}{2004}\natexlab{}.
\newblock \showarticletitle{Enforcing robust declassification}. In
  \bibinfo{booktitle}{\emph{Proceedings. 17th IEEE Computer Security
  Foundations Workshop, 2004.}} IEEE, \bibinfo{pages}{172--186}.
\newblock


\bibitem[Pullicino(2014)]%
        {DBLP:journals/corr/Pullicino14}
\bibfield{author}{\bibinfo{person}{Kyle Pullicino}.}
  \bibinfo{year}{2014}\natexlab{}.
\newblock \showarticletitle{Jif: Language-based Information-flow Security in
  Java}.
\newblock \bibinfo{journal}{\emph{CoRR}}  \bibinfo{volume}{abs/1412.8639}
  (\bibinfo{year}{2014}).
\newblock
\showeprint[arXiv]{1412.8639}
\urldef\tempurl%
\url{http://arxiv.org/abs/1412.8639}
\showURL{%
\tempurl}


\bibitem[Simonet(2003)]%
        {simonet2003flow}
\bibfield{author}{\bibinfo{person}{Vincent Simonet}.}
  \bibinfo{year}{2003}\natexlab{}.
\newblock \showarticletitle{The flow caml system}.
\newblock \bibinfo{journal}{\emph{Software release. Located at http://cristal.
  inria. fr/\~{} simonet/soft/flowcaml}}  \bibinfo{volume}{116}
  (\bibinfo{year}{2003}), \bibinfo{pages}{119--156}.
\newblock


\bibitem[Warren et~al\mbox{.}(2016)]%
        {warren2016cybercrime}
\bibfield{author}{\bibinfo{person}{Tucker Warren}, \bibinfo{person}{Jared
  Favole}, \bibinfo{person}{Scott Haber}, {and} \bibinfo{person}{Emily
  Hamilton}.} \bibinfo{year}{2016}\natexlab{}.
\newblock \showarticletitle{Cybercrime costs more than you think}.
\newblock \bibinfo{journal}{\emph{Hamilton Place Strategies Report}}
  (\bibinfo{year}{2016}).
\newblock


\bibitem[Weiss et~al\mbox{.}(2019)]%
        {DBLP:journals/corr/abs-1903-00982}
\bibfield{author}{\bibinfo{person}{Aaron Weiss}, \bibinfo{person}{Daniel
  Patterson}, \bibinfo{person}{Nicholas~D. Matsakis}, {and}
  \bibinfo{person}{Amal Ahmed}.} \bibinfo{year}{2019}\natexlab{}.
\newblock \showarticletitle{Oxide: The Essence of Rust}.
\newblock \bibinfo{journal}{\emph{CoRR}}  \bibinfo{volume}{abs/1903.00982}
  (\bibinfo{year}{2019}).
\newblock
\showeprint[arXiv]{1903.00982}
\urldef\tempurl%
\url{http://arxiv.org/abs/1903.00982}
\showURL{%
\tempurl}


\end{thebibliography}

%% Appendix
\appendix

\section{Appendix}

\subsection{Formal System: Extending Rust}\label{sec:formalizing}

We now formally describe our IFC system as an extension to Rust's type system.
The formal model of Rust is provided by Oxide
~\cite{DBLP:journals/corr/abs-1903-00982}. We take inspiration from the work on
tracking information flow in Rust presented by Flowistry
~\cite{10.1145/3519939.3523445}, though we diverge significantly from it in
order to capture the full expressivity of our system. Rust was chosen as the
foundation for our system because, as noted in~\cite{10.1145/3519939.3523445},
the language already performs a highly precise pointer analysis as part of
borrow checking, which is essential for any information flow system. Users need
not contend with an overly conservative pointer analysis, or with any
additional restrictions imposed by a precise one. Additionally, there is very
little existing work on IFC in Rust, and given the language's history of
embracing lightweight correctness techniques, we feel that our approach is
likelier to succeed within it. Our extensions to Oxide are colored in
\textcolor{brown}{brown}. We only consider flows between variables here,
because more complicated flow declarations such as those between functions and
modules do not present additional challenges to the formal system, behaving
similarly to the constructs already handled by it.

\begin{figure}[h]
    \begin{math}
        \begin{matrix}
            \text{Variables}~x & \text{Naturals}~m,n,k &
            \text{Concrete Regions}~r & \text{Type Vars.}~\alpha \end{matrix}
    \end{math}\\
    \begin{minipage}{0.5\textwidth}
    \begin{align*}
        \text{Path}~q~                      & ::=~\epsilon~|~n.q                \\
        \text{Places}~\pi~                  & ::=~x.q                           \\
        \text{Place Exprs.}~p~              & ::=~x~|*p~|~p.n                   \\
        \new{\text{Dependency Exprs.}~d~}   & ::=~\new{\delta~|~p}              \\
        \new{\text{Access Exprs.}~a~}       & ::=~\new{*~|~p}                   \\
        \text{Place Expr. Contexts}~p^\Box~ & ::=~\Box~|~*p^\Box~|~p^\Box.n
    \end{align*}
    \end{minipage}\hfill
    \begin{minipage}{0.5\textwidth}
        \begin{align*}
        \text{Loans}~l~                     & ::=~^\omega p                     \\
        \text{Ownership Qualifiers}~\omega~ & ::=~\texttt{shrd}~|~\texttt{uniq} \\
        \text{Constants}~c~                 & ::=~()~|~n~|~\texttt{true}~|~
            \ldots                                                              \\
        \text{Expressions}~e~               & ::=~c~|~p~|~\&r \omega p~|~\ldots \\
        \text{Base Types}~\tau^\text{B}~    & ::=~\texttt{u32}~|~
            \texttt{bool}~|~\ldots                                              \\
        \text{Sized Types}~\tau^\sisc~      & ::=~\tau^\text{B}~|~\alpha~|~
            \ldots
        \end{align*}
    \end{minipage}
    \caption{Oxide's syntax, extended}\label{fig:formal_synta}
\end{figure}

Beginning with the syntax, we present the parts of Oxide's syntax most relevant
to our discussion. We make two additions, the first of which introduces
\textit{dependency expressions} $d$, combining $\delta$, which tracks
dependencies, and place expressions $p$, which are used by Oxide to represent
values which can be assigned to. The other addition is similar, creating
\textit{access expressions} $a$, combining \texttt{*}, which appears in
Fig.~\ref{fig:case_study_rocket}, with place expressions.

Our typing judgement is 
$\Sigma; \Delta; \Gamma; \Theta; \new{\Pi; \Psi; \Xi} \vdash{} e:
\tau{} \new{~\bullet~\delta} \Rightarrow{} \Gamma'\new{; \Pi'}$,
which reads ``given the global
environment $\Sigma$, the type environment $\Delta$, the stack typing
environment $\Gamma$, the continuation typing environment $\Theta$, the
dependency environment $\Pi$, the flow policy environment $\Psi$, and the
branch dependency environment $\Xi$, the expression $e$ with type $\tau$ and
dependencies $\delta$ produces an updated stack typing environment $\Gamma'$
and dependency environment $\Pi'$''. This is similar to the one used by
Flowistry, with additional environments for tracking flow policies and control
flow dependencies--- the latter with increased precision.

\begin{figure}
    \begin{small}
        % rule for booleans
        \textsc{T-True}\hspace{19em}
        \textsc{T-Move}\\
        \vspace{1em}
        \typingrule{}{0}{}
        { \Sigma; \Delta; \Gamma; \Theta; \new{\Pi; \Psi; \Xi} }
        { \texttt{true}: \texttt{bool}
            \new{~\bullet~\varnothing} }
        { \Gamma; \new{\Pi} }\qquad
        % rule for move
        \typingrule{}{1}{\hypo{\begin{matrix}
                    \Delta; \Gamma; \Theta{} \vdash_{\texttt{uniq}} \pi{} \Rightarrow{}
                    \curlybr{^\texttt{uniq} \pi{}} \qquad
                    \Gamma(\pi) = \tau^\sisc{} \\
                    \texttt{noncopyable}_\Sigma{}~\tau^\sisc{}\qquad
                    \new{\delta{} = \deltaplace(\pi: \tau^\sisc, \Pi)}
                \end{matrix}}}
        { \Sigma; \Delta; \Gamma; \Theta; \new{\Pi; \Psi; \Xi} }
        { \pi: \tau^\sisc{} \new{~\bullet~\delta} }
        { \Gamma[\pi{} \mapsto{} \tau^{\sisc^\dagger}]; \new{\Pi} }\\
        \vspace{1em}
        % rule for copy
        \typingrule{T-Copy}{1}{\hypo{\begin{matrix}
                    \Delta; \Gamma; \Theta{} \vdash_\texttt{shrd} p \Rightarrow{}
                    \curlybr{\overline{l}}\qquad
                    \Delta; \Gamma{} \vdash_\texttt{shrd} p: \tau^\sisc       \\
                    \texttt{copyable}_\Sigma{}~\tau^\sisc{}\qquad
                    \new{\delta{} =
                        % we don't add Xi here because we only do that for exprs after the
                        % whole branch expr
                        \sideset{_{\tau^\sisc}^\varnothing}{}\bigsqcup_{^\omega{} p' \in{}
                    \curlybr{\overline{l}}} \deltaplace(p': \tau^\sisc, \Pi)} \\
                    % p should be included in its own loan set if it's accessible, so if
                    % p isn't a (de)reference here, it's okay!
                    % \new{$\Omega_1 =
                    %     \Omega \doublecup^+
                    %         \bigdoublecup_{^\omega p' \in \curlybr{\overline{l}}}^+
                    %         \relrules(p': \tau^\sisc, \Delta, \Gamma, \Pi, \Psi)$}
                \end{matrix}}}
        { \Sigma; \Delta; \Gamma; \Theta; \new{\Pi; \Psi; \Xi} }
        { p: \tau^\sisc{} \new{~\bullet~\delta} }
        { \Gamma; \new{\Pi} }\\
        \vspace{1em}
        \typingrule{T-Tuple}{1}{\hypo{\begin{matrix}
                    \forall{} i \in{} \{~1, \ldots, n~\}~.~ % chktex 39
                    \typejgmt[\Sigma; \Delta; \Gamma_{i-1};
                        \Theta, \tau_1^\sisc, \ldots, \tau_{i-1}^\sisc{};
                        \new{\Pi_{i-1}; \Psi; \Xi}]
                    {e_i: \tau_i^\sisc{} \new{~\bullet~\delta_i}}
                    {\Gamma_i; \new{\Pi_i}} \\
                    % We have to add the tuple's own label to the labels of its contents
                    % since a tuple is itself a base type!
                    % \new{$\delta = \bigcup_{i \in \curlybr{1, \ldots, n}}
                    %     \delta_i$}\qquad
                    % TODO: is this okay?? we need this for overriding on sources to work
                    % like we want, pretty sure!
                    \new{\forall{} i \in{} \curlybr{1, \ldots, n}~.~ % chktex 39
                        \forall{} p^\Box[\delta_i] \in{}
                        \leaves(\delta_i: \tau_i^\sisc)~.~ % chktex 39
                        p^\Box[\delta.i] = p^\Box[\delta_i]}
                \end{matrix}}}
        { \Sigma; \Delta; \Gamma_0; \Theta; \new{\Pi_0; \Psi; \Xi} }
        { (e_1, \ldots, e_n): (\tau_1^\sisc, \ldots, \tau_n^\sisc)
            \new{~\bullet~\delta} }
        { \Gamma_n; \new{\Pi_n} }
    \end{small}
    \caption{A collection of basic typing rules}\label{fig:simple_rules}
\end{figure}

\subsubsection{Tracking flows}

Our first rule, for the value \texttt{true}, appears in
Fig.~\ref{fig:simple_rules}. Bare values are effectively ignored by the
information flow system, producing an empty dependency expression $\delta$,
because flow rules cannot be declared on them. The next two rules,
\textsc{T-Move} and \textsc{T-Copy}, are more interesting, but require the
introduction of a few metafunctions. Much of the theory presented here is
founded on the treatment of places, $p$, as nothing more than the composition
of their \textit{leaves}. That is, given a place which represents a struct with
fields $f$ and $g$, where $g$ is itself a struct with field $h$, the leaves of
$p$ are $f$ and $h$. \textit{Dependencies} are always \textit{leaves}---
dependency sets contain the leaves of a place that is depended on, rather than
the `top level' place.
The $\deltaplace{}\!$ and $\delta_1
    \prescript{\curlybr{\!\overline{p}\!\!}}{\tau^\sisc}\sqcup~\delta_2$ operations
used by \textsc{T-Move} and \textsc{T-Copy} depend on this notion of analyzing
a place expression for the set of atomic places it contains; doing so allows
the analysis to be maximally precise without sacrificing simplicity. We exclude
type constructors, supported through the special inclusion of the
\either{} type by Oxide, because their treatment is similar to that for
fields. Oxide also does not support the declaration of recursive types, but
these can soundly be treated as leaves themselves in a practical
implementation. The definition of these two metafunctions is shown on the first
two lines of Fig.~\ref{fig:formal_metafunctions}.

\begin{figure}[h]
    \begin{math}
        \deltaplace(p: \tau^\sisc, \Pi) =
        \forall p_\mathrm{leaf}^\Box[p] \in \leaves(p: \tau^\sisc)~.~
        p_\mathrm{leaf}^\Box[\delta] =
        \curlybr{p_\mathrm{leaf}^\Box[p]} \cup
        \Pi(p_\mathrm{leaf}^\Box[p]); \delta
    \end{math}\\
    \vspace{1em}
    \begin{math}
        \delta_1
        \prescript{\curlybr{\!\overline{p}\!}}{\tau^\sisc}\sqcup~
        \delta_2 =
        \forall p_\mathrm{leaf}^\Box[\delta_1] \in
        \leaves(\delta_1: \tau^\sisc)~.~
        p_\mathrm{leaf}^\Box[\delta_r] =
        p_\mathrm{leaf}^\Box[\delta_1] \cup
        p_\mathrm{leaf}^\Box[\delta_2] \cup
        \curlybr{\overline{p}}; \delta_r
    \end{math}\\
    \vspace{1em}
    \begin{math}
        \assigndeps(p: \tau^\sisc, \delta, \curlybr{\overline{p}}) =
        \forall{} p_\mathrm{leaf}^\Box[p] \in{} \leaves(p: \tau^\sisc)~.~ % chktex 39
        p_\mathrm{leaf}^\Box[p] \mapsto{} p_\mathrm{leaf}^\Box[\delta]
        \cup{} \curlybr{\overline{p}}
    \end{math}\\
    \vspace{1em}
    % this works because the leaves of \delta should be dependency sets!
    \begin{math}
        \deltaleaves(\delta: \tau^\sisc) = \bigcup \leaves(\delta: \tau^\sisc)
    \end{math}
    \caption{Selected metafunctions for manipulating dependencies}\label{fig:formal_metafunctions}
\end{figure}

We elide the simple definition of the $\leaves$ metafunction, which has a case
for each of the syntactic forms an Oxide type might take. $\deltaplace$
generates a $\delta$ which contains all the dependencies of a place, retrieved
from $\Pi$. Note that $\delta$ mirrors the structure of the type of the place
whose dependencies it models, containing the same fields (specifically, leaves)
as the type it corresponds with. Each of these leaves of $\delta$ contains a
set of dependencies. It is treated as a place itself as a syntactic
simplification; it should be thought of as a map which is compatible with place
expression contexts $p^\Box$. The next metafunction, $\delta_1
    \prescript{\curlybr{\!\overline{p}\!\!}}{\tau^\sisc}\sqcup~\delta_2$, acts as a
union on a $\delta_1$ and a $\delta_2$ which correspond to the same type,
merging dependency sets between like leaves. An additional parameter,
$\curlybr{\overline{p}}$, can be used to unconditionally add dependencies to
all leaves of the resulting $\delta$.

\textsc{T-Move}, then, generates a $\delta$ which represents the dependencies
of a place $\pi$, using it to represent the dependencies of the moved variable.
\textsc{T-Copy} functions analogously, taking into account that \textit{place
expressions} (which may hold dereferences, see Fig.~\ref{fig:formal_synta}),
not just places, may be copied. As a result, it must use the ownership
judgement (in the first premise) to retrieve the dependencies of all the loans
for the place expression $p$ to be copied. A loan is any place expression which
represents the same abstract memory location as the place expression $p$, such
as a dereference of a shared reference to $p$. We use the $\sqcup$ operator to
merge each loan's generated $\delta$ with the rest. Finally, \textsc{T-Tuple}
demonstrates how $\delta$ is used with place expression contexts $p^\Box$ in
tracking dependencies.

\subsubsection{Preventing flows}

\begin{figure}
    % for binding!
    \typingrule{T-Let}{1}{
        \hypo{\begin{matrix}
                % type checking e1
                \typejgmt{e_1: \tau_1^\sisc{} \new{~\bullet~\delta} }
                % Remember that we don't explicitly say what the new flow rules
                % are here, we just get them as a result like delta!
                { \Gamma_1; \new{\Pi_1} }\qquad
                % rewriting e1's type to the type of x, the variable being bound
                \Delta; \Gamma_1; \Theta{} \vdash^+ \tau_1^\mathrm{SI}
                \leadsto{} \tau_a^\mathrm{SI} \rvdash{} \Gamma_1' \\
                % checking that none of the regions in the type of x have been
                % reborrowed (for ownership safety?)
                \forall{} r \in{} \freeregions(\tau_a^\mathrm{SI})~.~ % chktex 39
                \Gamma_1' \vdash{} r~\texttt{rnrb}\qquad
                % for the case where a_1 ->! *, which is the only kind of rule
                % which can be violated by let binding
                %
                % * ->! p_1 can't be violated by let bindings, since x isn't
                % flowing anywhere, and it hasn't been bound yet!
                %
                % the second argument could be any place
                \new{\allowed(\deltaleaves(\delta: \tau^\sisc) \cup{} \Xi,
                \curlybr{x: \tau_a^\sisc}, \Psi)}                 \\
                % update the dependencies for x in Pi
                \new{\Pi_1' =
                \Pi_1[\assigndeps(x: \tau_a^\sisc, \delta, \Xi)]} \\
                % type check the thing after the binding form
                \typejgmt[\Sigma; \Delta;
                    \gcloans_\Theta(\Gamma_1', x: \tau_a^\sisc);
                    \Theta; \new{\Pi_1'; \Psi; \Xi}]
                { e_2: \tau_2^\sisc{} \new{~\bullet~\delta'} }
                { \Gamma_2, x: \tau^\sdsc; \new{\Pi_2} }
            \end{matrix}}}
    { \Sigma; \Delta; \Gamma; \Theta;
        \new{\Pi; \Psi; \Xi} }
    { \texttt{let}~x: \tau_a^\sisc{} = e_1; e_2: \tau_2^\sisc{}
        \new{~\bullet~\delta'} }
    { \Gamma_2; \new{\Pi_2} }\\
    \vspace{1em}
    \typingrule{T-Assign}{1}{
    \hypo{\begin{matrix}
            % type checking the thing we're assigning (not assig to)
            \typejgmt{e: \tau^\sisc{} \new{~\bullet~\delta}}
            {\Gamma_1; \new{\Pi_1}}\qquad
            % getting the type of the thing we're assigning to
            % (might be dead, not a problem since we're about to reassign it)
            \Gamma_1 (\pi) = \tau^\sxsc{}                          \\
            % for Oxide proofs
            \tau^\sxsc{} = \& r~\omega~\tau^\xisc{} \Longrightarrow{}
            r~\mathrm{is~unique~to}~\pi{}~\mathrm{in}~\Gamma_1{}\qquad
            % Invalidate all dereferences of the place being assigned to
            \Delta; \Gamma_1{} \righttrianglebar{}\! *\pi; \Theta{} \vdash^=
            \tau^\sisc{} \leadsto{} \tau^\sxsc{} \rvdash{} \Gamma' \\
            % either it's definitely dead (moved?) or we have access to it?
            (\tau^\sxsc{} = \tau^\sdsc{} \lor{} \Delta; \Gamma'; \Theta{}
            \vdash_\uniq{} \pi{} \Rightarrow{}
            \curlybr{^\uniq{} \pi})\qquad
            % add the dependencies of pi to all the places that conflict with
            % it
            \new{\Pi_2 =
            \Pi_1[\assigndeps(\pi: \tau^\sisc, \delta, \Xi)]}      \\
            % error if this flow is not allowed
            \new{\forall{} p^\Box[\pi]: \tau_\mathrm{leaf}^\sisc{}
                \in{} \leaves(\pi: \tau^\sisc)~.
                \allowed(p^\Box[\delta] \cup{} \Xi,
                \curlybr{p^\Box[\pi]: \tau_\mathrm{leaf}^\sisc},
            \Psi)}                                                 \\
        \end{matrix}}}
    { \Sigma; \Delta; \Gamma; \Theta; \new{\Pi; \Psi; \Xi} }
    { \pi{}:= e: \texttt{unit} \new{~\bullet~\varnothing} }
    { \Gamma'[\pi{} \mapsto{} \tau^\sisc], \new{\Pi_2} }
    \caption{Selected rules responsible for checking flows}\label{fig:let_assig}
\end{figure}

A second selection of rules, presented in Fig.~\ref{fig:let_assig}, is
responsible not only for \textit{tracking} flows, but for \textit{preventing}
them when needed. The first is \textsc{T-Let}, which provides the base case for
information flow--- all dependencies tracked throughout the system originate
here. In order to explain \textsc{T-Let}, we introduce our final set of
metafunctions. $\assigndeps$ takes a place $p$, its $\delta$, and a set of
dependencies as arguments, and adds those dependencies to each leaf of $p$; it
is used to update a place's entry in $\Pi$. $\deltaleaves$ gathers all the
dependencies from a $\delta$, unioning the dependency sets from its leaves.
Finally, $\allowed$ (shown in Fig.~\ref{fig:formal_def_allowed}) takes a set of
sources, a set of destinations, and a flow policy set $\Psi$ as arguments, and
checks that each of the sources is allowed to flow to each of the destinations,
succeeding if all of the flows are allowed. It relies on $\getperms$ (shown
in Fig.~\ref{fig:formal_def_getperms}) to search $\Psi$ for the correct flow policy
to apply, which it does according to the overriding policy discussed in
Fig.~\ref{sec:motivation}. To do this, it relies on $\plustriangle$, a reflexive
and transitive binary relation on access expressions $a$ which will succeed if
the first access expression \textit{syntactically represents access} to the
second one. $p~\plustriangle p.f$ and $*~\plustriangle p$ define success.

\begin{figure}[h]
    \begin{minipage}{0.49\textwidth}
    \begin{align*}
        &\allowed(\curlybr{\overline{p_1: \tau_1^\sisc}},
            \curlybr{\overline{p_2: \tau_2^\sisc}}, \Pi, \Psi) =\\
            &\quad\forall (p_1: \tau_1^\sisc, p_2: \tau_2^\sisc) \in{}
                \curlybr{\overline{p_1: \tau_1^\sisc}} \times{}
                \curlybr{\overline{p_2: \tau_2^\sisc}}~.\\
            &\quad((\_, \_, \texttt{true}) = \getperms(p_1, p_2, \Psi))
    \end{align*}
    \caption{Checks if flows are allowed}\label{fig:formal_def_allowed}
    \end{minipage}\hfill
    \begin{minipage}{0.48\textwidth}
    \begin{align*}
        &\getperms(p_1,p_2,\Psi) =\\
        &\quad\!\foldr(((a_2, a_3, \beta_1), (a_4, a_5, \beta_2)) \rightarrow\\
        &\quad\quad\text{if}~a_2~\plustriangle{} p_1 \wedge{} a_3~\plustriangle p_2~\wedge\\
        &\quad\quad\quad((a_4~\plustriangle{} a_2 \wedge{} a_5~\plustriangle{} a_3)~\lor\\
        &\quad\quad\quad((a_4~\plustriangle{} a_2 \lor{} a_5~\plustriangle{} a_3) \wedge{}
                            \beta_1 = \texttt{false}))\\
        &\quad\quad\text{then}~(a_2, a_3, \beta_1)\\
        &\quad\quad\text{else}~(a_4, a_5, \beta_2),\\
        &\quad(*, *, \texttt{true}), \Psi)
    \end{align*}
        \caption{Retrieves a maximally precise rule from $\Psi$}\label{fig:formal_def_getperms}
    \end{minipage}
\end{figure}

Given these definitions, \textsc{T-Let} checks if the dependencies of the
expresssion $e$ to be bound, as well as any branch dependencies, are permitted
to flow to the variable to be bound, according to the current flow policy.
Additionally, it gives the bound variable the expression's dependencies by
updating $\Pi$. Similarly, \textsc{T-Assign} updates the assignee's
dependencies in the dependency environment with those of the expression being
assigned, along with any dependenies from control flow. It then checks whether
flows are allowed from each leaf of the assignee from the relevant dependencies
of the expression being assigned.

We now briefly discuss our handling of function application, eliding typing
rules for brevity. Note that we change Oxide's function call form, removing the
return value--- any program which uses function return values may
be transformed to a program which returns information through a mutable
reference passed to the function. This simplifies information flow checking for
function calls. Like~\cite{10.1145/3519939.3523445}, our system is
\textit{modular}, opting to preserve abstraction at the cost of precision. When
a function call is encountered, we first retrieve the flow policies declared by
the function which are valid for its entire scope, in addition to any policies
\textit{permitting} flows, adding them all to a new $\Psi$ representing the
function's flow policy set. We then substitute the function call's actual
parameters for its formal parameters within this $\Psi$, and compare it to the
$\Psi$ for the calling context. If the calling context's $\Psi$ implies the one
generated for the function call, then no error is raised. Specifically we check
that, for all the \textit{mutable references} passed to the function, any flow
permitted by the function to that unique reference is also permitted by the
calling context. We exclusively consider \texttt{uniq} refs because they are
the only means by which information may escape our function call form, and
violate the calling context's information flow policies.

The typing rules for the \flow\!\!\! and \allow{} constructs are
straightforward, so we also elide their definitions for brevity. When a flow
policy is declared, it is added to the current flow policy environment $\Psi$.
Flow policies may be declared between access expressions; for any access
expression which is not \texttt{*}, the ownership judgement is used to ensure
that the place expression is accessible. When \allow{} is used on an expression
$e$ of type $\tau$, it returns a $\delta$ with all of its leaves (according to
$\tau$) set to $\varnothing{}$.

\end{document}